\documentclass[aps,prl,twocolumn,preprintnumbers,superscriptaddress,
            nofootinbib]{revtex4}

\newcommand{\PRE}[1]{}				

\usepackage{amsmath}
\usepackage{amsfonts}
\usepackage{amssymb}
\usepackage{epsfig}
\usepackage{hyperref}
\usepackage{graphicx}

\def\myK{{\cal K}}

\begin{document}

\title{\PRE{\vspace*{1.5in}}
A Measurement of Cubic-Order Primordial Non-Gaussianity ($g_{\rm NL}$ and $\tau_{\rm NL}$) \\ With WMAP 5-Year Data
\PRE{\vspace*{0.3in}}}

\author{Joseph Smidt\footnote{ jsmidt@uci.edu}}
\affiliation{Center for Cosmology, Department of Physics and Astronomy,
University of California, Irvine, CA 92697, USA }

\author{Alexandre Amblard} 
\affiliation{Center for Cosmology, Department of Physics and Astronomy,
University of California, Irvine, CA 92697, USA }

\author{Asantha Cooray} 
\affiliation{Center for Cosmology, Department of Physics and Astronomy,
University of California, Irvine, CA 92697, USA }

\author{Alan Heavens} 
\affiliation{Scottish Universities Physics Alliance (SUPA),~ Institute for Astronomy,
University of Edinburgh, Blackford Hill,  Edinburgh EH9 3HJ, UK}

\author{Dipak Munshi} 
\affiliation{Scottish Universities Physics Alliance (SUPA),~ Institute for Astronomy,
University of Edinburgh, Blackford Hill,  Edinburgh EH9 3HJ, UK}
\affiliation{School of Physics and Astronomy, Cardiff University, CF24 3AA}

\author{Paolo Serra}
\affiliation{Center for Cosmology, Department of Physics and Astronomy,
University of California, Irvine, CA 92697, USA }

\date{\today}

\begin{abstract}
\PRE{\vspace*{.3in}}

We measure two higher-order power spectra involving weighted cubic and
squared temperature anisotropy maps from WMAP 5-year data to study
the trispectrum generated by primordial non-Gaussianity. Using these measurements combined with Gaussian and noise simulations, we constrain  the cubic order
non-Gaussianity parameters $\tau_{\rm NL}$, and $g_{\rm NL}$.  With V+W-band data out to $l_{\rm max}=600$, we find  $-7.4 < g_{\rm NL}/10^5 < 8.2$ and   $-0.6 < \tau_{\rm NL}/10^4 < 3.3$ improving the previous COBE-based limit on $\tau_{\rm NL} < 10^8$ nearly four orders of magnitude with WMAP. We find that the ratio of trispectrum to bispectrum amplitude as
captured by the ratio of $\tau_{\rm Nl}/(6f_{\rm NL}/5)^2$ ranges from -3 to 21 at the 95\% confidence level.

\end{abstract}


\maketitle


{\it Introduction.}---Cosmic Inflation has deservedly become a cornerstone of modern cosmology~\cite{Guth:1980zm,Linde:1981mu,Albrecht:1982wi}. Inflation solves the flatness, horizon and the monopole problems of the standard Big-Bang cosmology. Furthermore, inflation is the prevailing paradigm related to the origin of density perturbations that gave rise to the large-scale structure we see in the universe. It posits that a nearly exponential expansion stretched space in the first moments of the early universe and promoted microscopic quantum fluctuations to perturbations on cosmological scales today~\cite{GuthPi,Bardeen}. 

In the simple scenario, inflation is driven by a single scalar field whose potential energy dominates it's kinetic energy.  This ``slow-roll'' situation leads to an exponential expansion of the cosmic spacetime via the Einstein equations coupled to a scalar field.  In order to maintain a slow roll, the scalar field must have minimal self interactions.  Such a non-interacting field has the statistical feature that its fluctuations are Gaussian.  A Gaussian field contains no correlations between points arising from interactions.

In addition to the standard inflationary scenario, several possible mechanisms of inflation have emerged. These models usually involve multiple fields and or other exotic objects such as branes (motivated by string theory) that have non-trivial interactions.  These interactions produce a departure from Gaussianity in a model-dependent manner~\cite{Byrnes:2010em,Engel:2008fu,Chen:2009bc,Boubekeur:2005fj}.  Constraining non-Gaussianity therefore is important to distinguish between the plethora of inflationary models~\cite{Komatsu:2009kd}. 

The first order non-Gaussian parameter, $f_{\rm NL}$, has been measured with increasing success in the bispectrum, or three-point correlation function of temperature anisotrpies of the cosmic microwave background (CMB).   Such studies have found $f_{\rm NL }$ to be consistent with zero~\cite{Yadav:2007yy,Smith:2009jr,Komatsu:2010fb,Smidt:2009ir}.  

In addition to $f_{\rm NL}$, with the trispectrum or four point correlation function of CMB anisotrpies, we can measure second order non-Gaussian parameters $\tau_{\rm NL}$ and $g_{\rm NL}$.  Constraints on $\tau_{\rm NL}$ have never been directly preformed with data.  The very weak constraint often quoted in the literature with $\tau_{\rm NL} < 10^8$ is an estimate based on a null detection of the COBE data~\cite{Boubekeur:2005fj,Kunz:2001ym,Komatsu:2002db}.  Furthermore, the ratio  between $\tau_{\rm NL}$ and $(6f_{\rm NL}/5)^2$, involving the
amplitudes of trispectrum and bispectrum, is often is often a constraint on inflationary models~\cite{Byrnes:2010em,Chen:2009bc}.  In fact, for many inflationary models this ratio is related to the
tensor-to-scalar ratio, which is now probed by CMB polarization data,
providing additional consistency relations to understand the
underlying physics of inflation~\cite{Byrnes:2010em}. Constraints on $g_{\rm NL}$ have only recently been preformed, but not with the trispectrum of the CMB~\cite{Vielva:2009jz}.    
 
In this analysis we use the trispectrum, or the four-point correlation function of temperature anisoptropies, to measure primordial non-Gaussianity at second order using WMAP 5-year data~\cite{Hu:2001fa}.  
 
{\it Theory.}--- 
 To parameterize the non-Gaussianity of a nearly Gaussian field, such as the primordial curvature perturbations $\Phi({\bf x})$,  we can expand it perturbatively~\cite{Kogo:2006kh} to second order as:
\begin{equation}
\Phi({\bf x}) = \phi_L({\bf x}) + f_{\rm NL} \left[\phi_L^2({\bf x}) - \langle \phi^2_L({\bf x})\rangle\right] + g_{\rm NL} \phi_L^3({\bf x}) 
\label{eq:phi}
\end{equation}
 where $\phi_L({\bf x})$ is the purely Gaussian part with $f_{\rm NL}$ and $g_{\rm NL}$ parametrizing  the first and second order deviations from Gaussianity.  Fortunately, information about the curvature perturbations are contained within the CMB through the spherical harmonic coefficients of the temperature anisotropies:
\begin{eqnarray}
a_{lm} &=& 4 \pi (-i)^l \int \frac{d^3{\bf k}}{(2 \pi)^3} \Phi({\bf k})  g_{Tl}(k)Y_{l}^{m*}({\hat{\bf k}}) \\ 
 \theta(\hat{\bf n}) &=& \frac{\Delta T}{T}(\hat{\bf n}) = \sum_{l m} a_{lm}Y_{l}^{m*}({\hat{\bf n}})\, ,
\label{eq:alm} 
\end{eqnarray}
where $\Phi({\bf k})$ are the primordial curvature perturbations, $g_{Tl}$ is the radiation transfer
function, $\theta$ is the field of temperature fluctuations in the CMB and $Y^{l}_m$'s are the spherical harmonics.

If the curvature perturbations are purely Gaussian, all the statistical information we can say about them is contained in the two point correlation function $\left<\Phi({\bf x_1}) \Phi({\bf x_2})\right> $. The information contained in the two point function is usually extracted in spherical harmonic space, leading to the power spectrum $C_l$, defined by: 
\begin{eqnarray}
C_l = \left<a_{lm} a_{lm}\right> = {1 \over (2l+1)}\sum_m a_{lm} a^*_{lm}
\label{eq:powerspec}
\end{eqnarray}
However, if the curvature perturbations are slightly non-Gaussian, this two point function is no longer sufficient to articulate all the information contained in the field.  With non-Gaussianity, extra information can be extracted from the three, four and higher n-point correlation functions.  

In this paper we look to the trispectrum, or four point correlation function, to probe non-Gaussianity.  To extract information from the four point function $\left<\Phi({\bf x_1}) \Phi({\bf x_2})\Phi({\bf x_3}) \Phi({\bf x_4})\right> $, we again work in spherical harmonic space and compute:
\begin{eqnarray}
\label{eq:tripieces}
\left<a_{l_1m_1} a_{l_2m_2} a_{l_3m_3} a_{l_4m_4}\right> &=& \\
\left<a_{l_1m_1} a_{l_2m_2} a_{l_3m_3} a_{l_4m_4}\right>_G\,   
&+& \left<a_{l_1m_1} a_{l_2m_2} a_{l_3m_3} a_{l_4m_4}\right>_c.\nonumber
\end{eqnarray}
where we see that the four point function breaks up into a piece representing a Gaussian contribution and a second piece representing the non-Gaussian contribution.  The non-Gaussian piece is commonly referred to as the connected piece and is generated by interactions between points.

\begin{figure}[t]
   \vspace{-0.2cm}
    \begin{center}
      \includegraphics[scale=0.45]{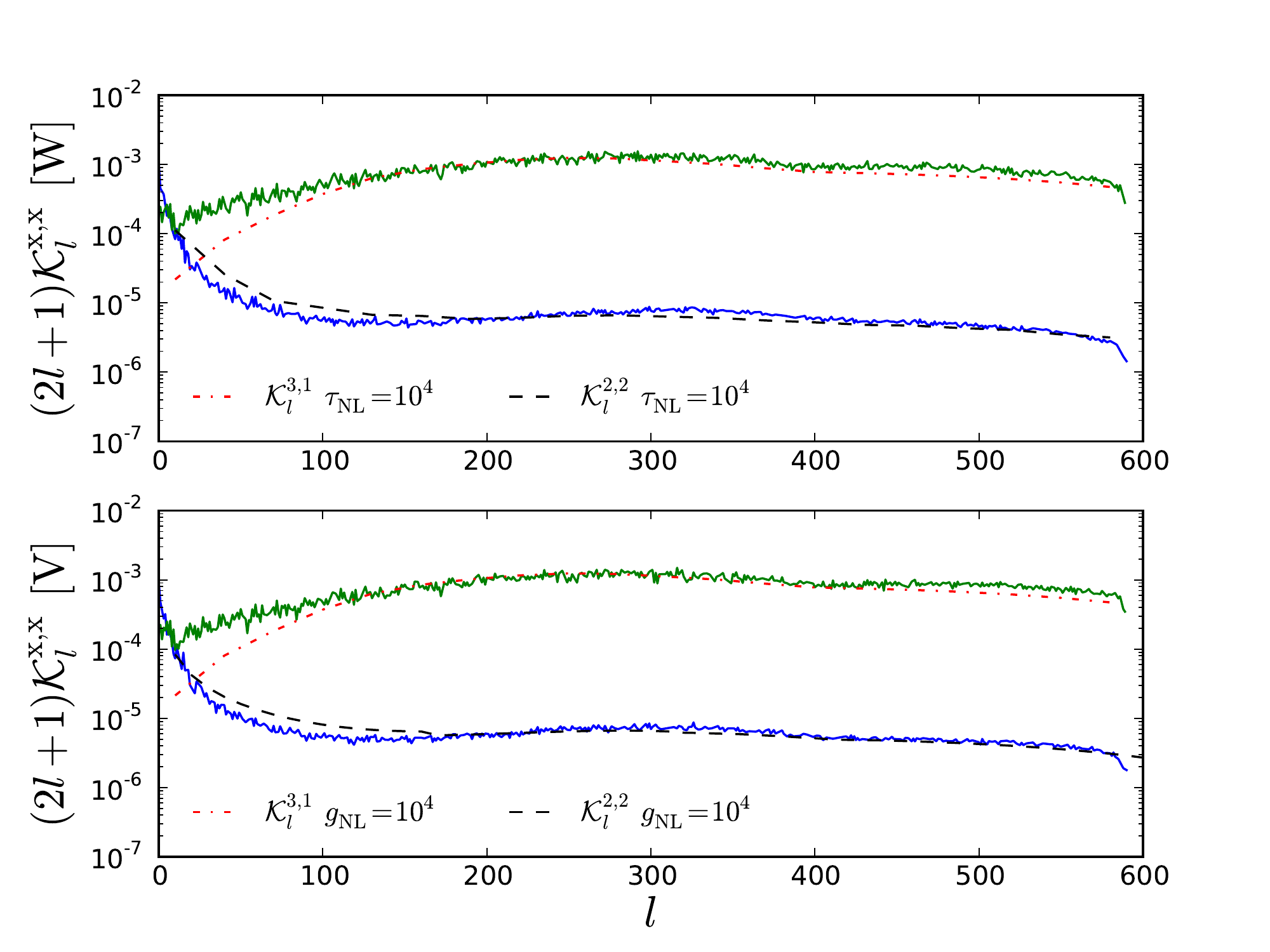} 
   \end{center}
   \vspace{-0.7cm}
   \caption[width=3in]{The top plot shows the ${\cal K}_l^{3,1}$ and ${\cal K}_l^{2,2}$ estimators, shown in green and blue respectively, taken from data for the W band.  The same estimators for the V band are shown on the bottom.  Additionally on the top the theoretical contributions for  ${\cal K}_l^{2,2}$ and ${\cal K}_l^{3,1}$ proportional to $\tau_{\rm NL}$ are shown with the bottom showing those proportional to $g_{\rm NL}$.  The Gaussian contributions were not removed from these plots.\vspace{-0.1cm}}
   \label{fig:wtheory}
\end{figure}

As with the two point function, from the four point function we can create various power spectra.  Two independent spectra that can be produced are ${\cal K}_l^{(3,1)} $ and ${\cal K}_l^{(2,2)}$  (See Fig.~\ref{fig:wtheory}) that are related to various compressions of $a_{lm}$s as follows:
\begin{equation}
 \label{eq:ak22e}
 {\myK}_l^{(2,2)} \sim  \left<a^{(2)}_{lm} a^{(2)*}_{lm}\right>, \hspace{0.5cm} {\myK}_l^{(3,1)} \sim   \left<a^{(3)}_{lm} a^{(1)*}_{lm}\right>. \\
\end{equation}
Here $\sim$ should be read as {\it is related to} as the full expression is very complex. (We refer the reader to Refs.~\cite{Munshi:2009wy}for full details of the
derivation of eq.~\ref{eq:ak22e} including the pre-factors.). The superscript in the $a^{({\rm x})}_{lm}$ refers to how many temperature maps must be combined to form the $a_{lm}$.  For example, only one is used in eq.~\ref{eq:alm}-\ref{eq:powerspec}.In the same manner $C_l$ is the angular power spectrum of temperature and ${\cal K}_l^{(2,2)}$ and ${\cal K}_l^{(3,1)}$ are the angular power spectrum of squared-temperature correlated against squared temperature and cubic
temperature correlated against temperature, respectively.  To optimize for primordial non-Gaussianity detection, however, the
temperature maps used for ${\cal K}_l^{(2,2)}$ and ${\cal K}_l^{(3,1)}$ are
weighted
appropriately in the same manner we optimized $C_l^{(2,1)}$ statistic for the
bispectrum measurement~\cite{Smidt:2009ir}.  

The estimators ${\cal K}_l^{(3,1)} $ and ${\cal K}_l^{(2,2)}$ contain contributions from the Gaussian and connected piece of equation~\ref{eq:tripieces}.  The connected piece further breaks up into two independent pieces parameterized by two parameters $\tau_{\rm NL}$ and $g_{\rm NL}$:
\begin{eqnarray}
 \label{eq:k22e}
 {\myK}_l^{(2,2)} |_c &=& \tau_{\rm NL} {\cal A}^{(2,2)}_l + g_{\rm NL}{\cal B}^{(2,2)}_l\\
  \label{eq:k31param}
 {\myK}_l^{(3,1)} |_c &=&   \tau_{\rm NL} {\cal A}^{(3,1)}_l  + g_{\rm NL}{\cal B}^{(3,1)}_l 
\end{eqnarray}
Here, ${\cal A}^{(\rm x,x)}_l$ and ${\cal B}^{(\rm x,x)}_l$ are the theoretical expectations for the connected part of  ${\cal K}_l^{(2,2)}$ and ${\cal K}_l^{(3,1)}$ with the assumption that $\tau_{\rm NL}$ and $g_{\rm NL}=1$ respectively.

It is clear from equation~\ref{eq:phi} that $g_{\rm NL}$ is a non-Gaussian parameter of second order.  Additionally, in a derivation of the trispectrum one finds that $\tau_{\rm NL}$ is related to $f^2_{\rm NL}$ in a model dependent way, and is therefore also a second order parameter.  

Since the exact relation between $\tau_{\rm NL}$ from the trispectrum and $f_{\rm NL}$ from the bispectrum is model dependent, this relationship gives another test to distinguish between models for primordial perturbations~\cite{Byrnes:2010em,Engel:2008fu,Chen:2009bc,Boubekeur:2005fj}.  It is common to compare $\tau_{\rm NL}$ with $(6f_{\rm NL}/5)^2$ and therefore  we constrain the relation $ A_{\rm NL} = \tau_{\rm NL}/(6f_{\rm NL}/5)^2$ where $A_{\rm NL}$ is the amplitude of   the trispectrum to the squared bispectrum. For many models $A_{\rm NL}$ can be used to further constrain the tensor-to-scalar ratio $r$~\cite{Byrnes:2010em}.

{\it Analysis and Results.}---Our recipe for analysis is
\begin{enumerate}
\item We calculate ${\cal A}^{(2,2)}_l$, ${\cal B}^{(2,2)}_l$, ${\cal A}^{(3,1)}_l$ and  ${\cal B}^{(3,1)}_l$ in Eq.~\ref{eq:k22e}-~\ref{eq:k31param} for $\tau_{\rm NL}$ and $g_{\rm NL} = 1$. (See~\cite{Munshi:2009wy} for more details on how to do this and other steps.)
\item We extract ${\cal K}_l^{(3,1)} $ and ${\cal K}_l^{(2,2)}$ directly from WMAP 5-year data.
\item We preform the extraction of ${\cal K}_l^{(3,1)} $ and ${\cal K}_l^{(2,2)}$ from 250 Gaussian maps, allowing us to determine error bars and the Gaussian piece of each estimator.
\item We subtract off the Gaussian contribution to these estimators to ensure we are fitting to the non-Gaussian contribution.
\item We fit the two unknowns $\tau_{\rm NL}$ and $g_{\rm NL}$ from data using the two equations simultaneously.  The amplitudes the theoretical curves must be scaled by gives the values for $\tau_{\rm NL}$ and $g_{\rm NL}$
\item We constrain $A_{\rm NL}$ by comparing $\tau_{\rm NL}$ from the trispectrum with $(6f_{\rm NL}/5)^2$ coming from the bispectrum.
\end{enumerate}
This recipe is described in grater detail below:

First we calculate ${\cal A}^{(2,2)}_l$, ${\cal B}^{(2,2)}_l$, ${\cal A}^{(3,1)}_l$ and  ${\cal B}^{(3,1)}_l$ theoretically using the full equations described in~\cite{Munshi:2009wy}.  To obtain $C_l$ we use CAMB~\cite{Lewis:1999bs}\footnote{${\rm http://camb.info/}$} with the WMAP 5-year best fit parameters and use the beam transfer functions from the WMAP team.    We then obtain the connected piece using a modified version of the CMBFAST code~\cite{Seljak:1996is}\footnote{${\rm http://www.cfa.harvard.edu/~mzaldarr/CMBFAST/cmbfast.html}$}. Plots of many of the quantities used for these calculations can be found in Ref.~\cite{Smidt:2009ir}. 

We combine ${\cal A}^{(\rm x,x)}_l$, ${\cal B}^{(\rm x,x)}_l$ with various assumed values of $\tau_{\rm NL}$ and $g_{\rm NL}$ to establish ${\cal K}_l^{(2,2)} $ and ${\cal K}_l^{(3,1)}$.  These are plotted in Figure~\ref{fig:wtheory}. These curves will be compared with estimators derived from data to determine the magnitude of each statistic.  Since we have two estimators, we can solve for the two unknowns $\tau_{\rm NL}$ and $g_{\rm NL}$ by fitting both estimators simultaneously.  

To calculate\footnote{see Smidt el al. 2009 for a similar calculation using the bispectrum for more details.~\cite{Smidt:2009ir}} the estimators from data, used in the lefthand side of equations~(\ref{eq:k22e}) and~(\ref{eq:k31param}), we use both the raw and foreground-cleaned WMAP 5-Year Stokes I maps for V- and W-bands masked with the KQ75 mask~\footnote{${\rm http://lambda.gsfc.nasa.gov/}$}. We use the Healpix library to analyze the maps.  For this analysis we only considered data out to $l_{\rm max} = 600$.  We correct for the KQ75 mask using a matrix $M_{l l'}$, based on the power spectrum of the mask, as described in~\cite{Munshi:2009wy}.

Figure~\ref{fig:wtheory} shows the results for ${\cal K}_l^{3,1}$ and ${\cal K}_l^{2,2}$ for the V and W frequency bands extracted from the raw WMAP 5-Year maps. 
In order to do proper statistics for our data fitting we create 250 simulated Gaussian maps of each frequency band with $n_{\rm side}=512$.  To obtain Gaussian maps we run the {\it synfast} routine of Healpix with an in-file representing the WMAP 5-year best-fit CMB anisotropy power spectrum and generate maps with information out to $l = 600$.  We then use {\it anafast}, without employing an iteration scheme, masking with the $KQ75$ mask, to produce $a_{lm}$'s for the Gaussian maps out to $l=600$.  Obtaining estimators from these Gaussian maps allows us to uncover the Gaussian contribution to each estimator in addition to providing us information needed to calculate the error bars on our results.  

This whole process is computationally intensive.  To calculate all theoretical estimators took nearly 8,000 CPU hours.  Furthermore, all the estimators from Gaussian and data maps combined took an additional 1600 CPU hours.
\begin{figure}
   \vspace{-0.2cm}
    \begin{center}
      \includegraphics[scale=0.45]{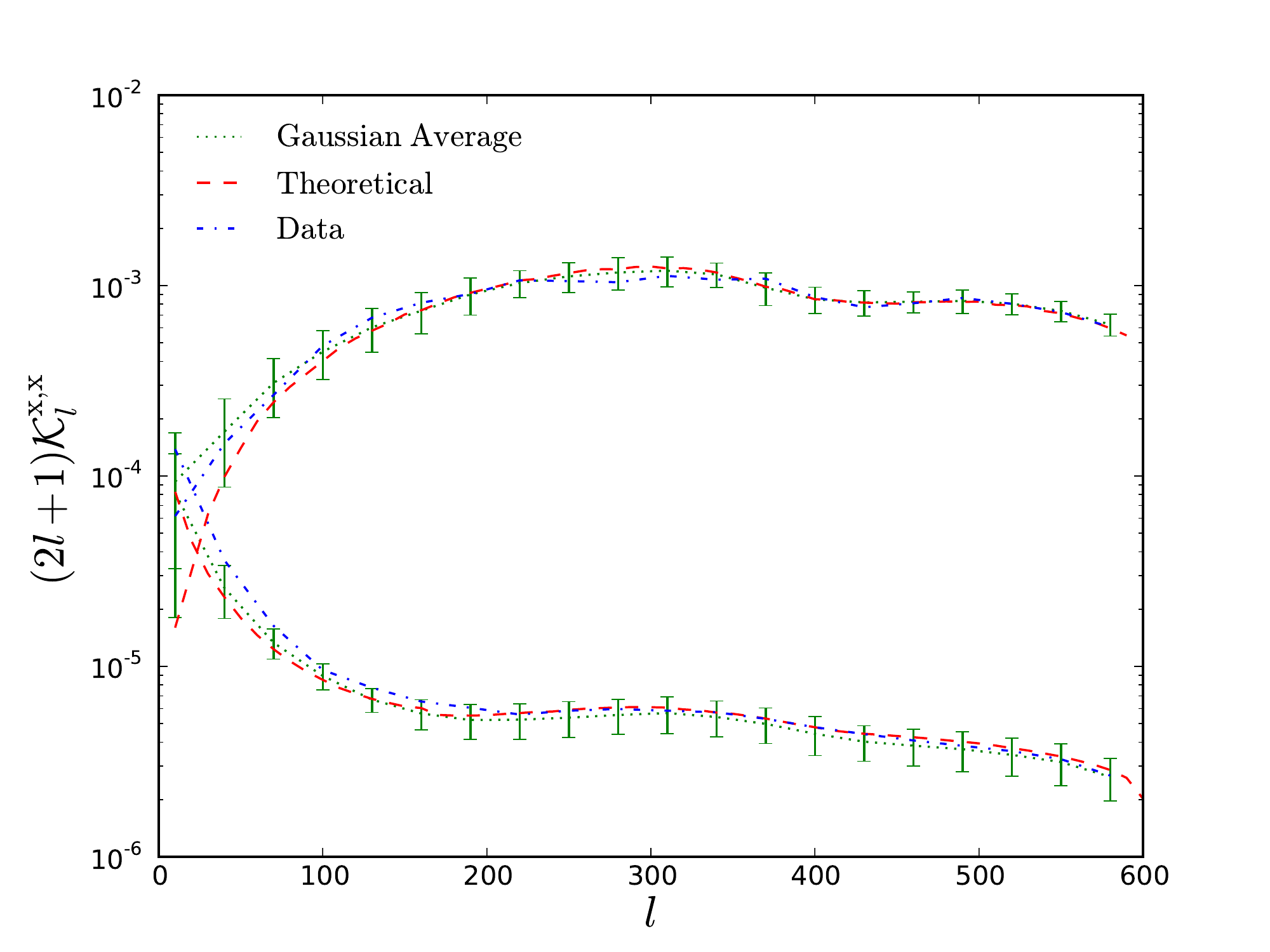} 
   \end{center}
   \vspace{-0.5cm}
   \caption[width=3in]{The relation between the full estimators coming from data versus the Gaussian contributions.  The green curve show the Gaussian contributions coming from averaging the estimators from the Gaussian maps.  The red curve is the theoretical Gaussian piece calculated from Eq.~\ref{eq:gausspiece} using the WMAP-5 best-fit cosmology power spectrum.  The error bars show two standard deviations from the Gaussian curves.  These curves are from W band data.\vspace{-0.1cm}}
   \label{fig:compare}
\end{figure}
\begin{figure}[t]
   \vspace{-0.2cm}
    \begin{center}
      \includegraphics[scale=0.48]{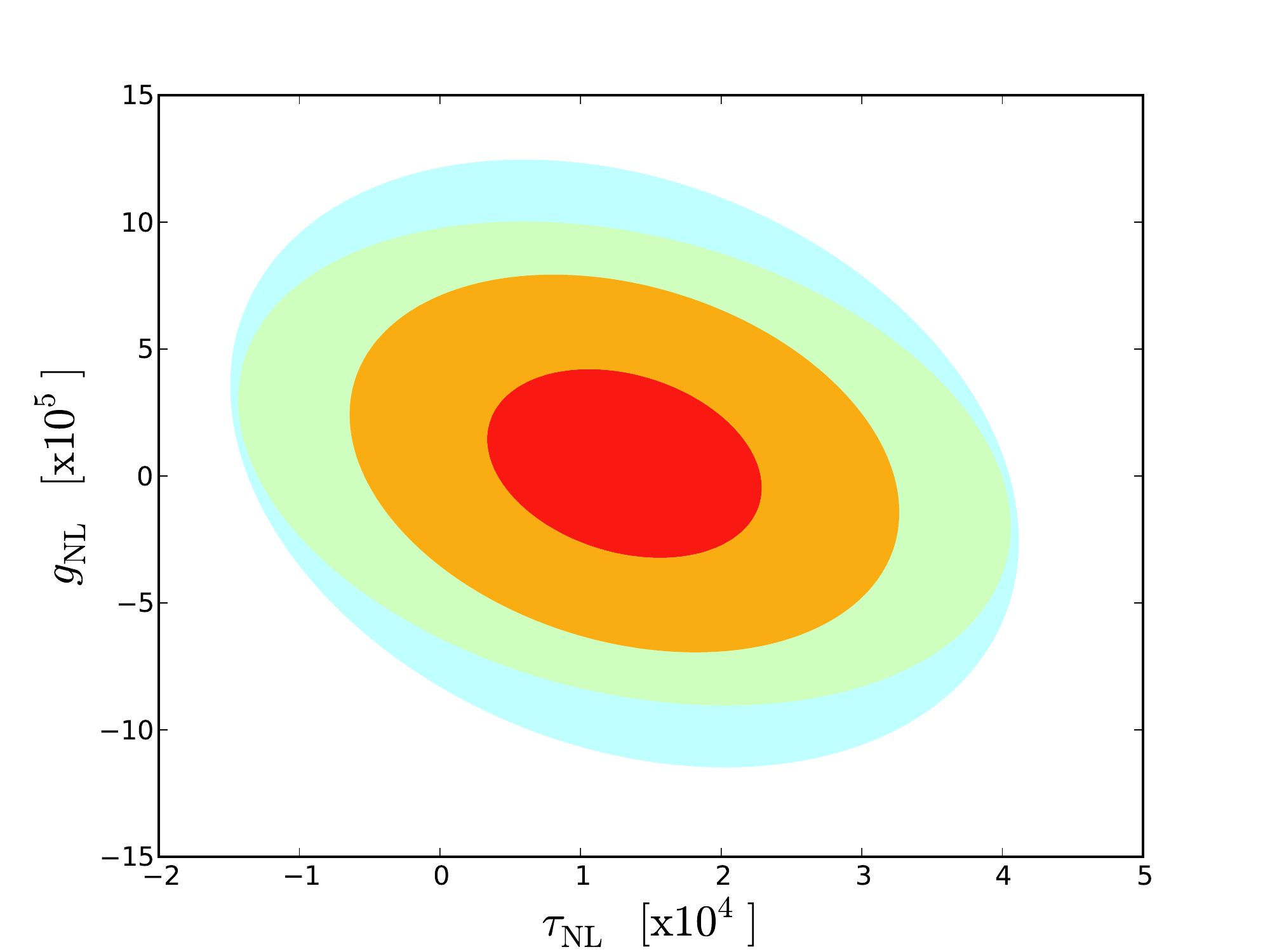} 
   \end{center}
   \vspace{-0.7cm}
   \caption[width=3in]{The 95\% confidence levels for $g_{\rm NL}$ versus $\tau_{\rm NL}$.  The red and orange represent the 68\% and 95\% intervals respectively for the combined V+W analysis.  The light blue regions represent the 95\% confidence intervals for the V band analysis, and the light green regions are for the W band.}
   \label{fig:confidence}
\end{figure}

As previously discussed, the full trispectrum can be decomposed into both a Gaussian and non-Gaussian or connected piece.  To make a measurement of  non-Gaussianity we to subtract off the Gaussian piece from the full trispectrum.  Figure~\ref{fig:compare} shows the the relationship between the full trispectrum and the Gaussian piece.  In this plot the Gaussian piece was calculated in two different ways as a sanity check.  First, the Gaussian maps were averaged over.  Second, the Gaussian piece of each estimator is calculated theoretically as described in\cite{Munshi:2009wy}.  
\begin{table}[t]
  \centering
  \begin{tabular}{@{} |c|c|c|c|@{}}
   \hline
   Band & W & V & V+W \\
   \hline
   Raw & & & \\
    $g_{\rm NL}$  & $4.7 {\rm x} 10^4 \pm 5.3  {\rm x} 10^5$ & $4.6 {\rm x} 10^4 \pm 5.9  {\rm x} 10^5$& $4.7 {\rm x} 10^4 \pm 3.9  {\rm x} 10^5$ \\ 
    $\tau_{\rm NL}$ & ${(1.63 \pm 1.27)}  {\rm x} 10^4$ &  ${(1.68 \pm 1.31)}  {\rm x} 10^4$ &  ${(1.64 \pm 0.98)}  {\rm x} 10^4$  \\ 
        $A_{\rm NL}$ &  $7.4 \pm 7.3$& $6.3 \pm 6.0$  &  $11.1 \pm 7.3$  \\ 

       \hline
       FC & & & \\
        $g_{\rm NL}$  & $4.2 {\rm x} 10^4 \pm 5.3  {\rm x} 10^5$ & $4.1 {\rm x} 10^4 \pm 5.9  {\rm x} 10^5$& $4.2 {\rm x} 10^4 \pm 3.9  {\rm x} 10^5$ \\ 
    $\tau_{\rm NL}$ &  ${(1.32 \pm 1.27)}  {\rm x} 10^4$ &   ${(1.39 \pm 1.31)}  {\rm x} 10^4$ &  ${(1.35 \pm 0.98)}  {\rm x} 10^4$  \\ 
            $A_{\rm NL}$ &  $6.0 \pm 6.7$& $5.2 \pm 5.7$  & $9.2 \pm 6.1$  \\ 

           \hline
    \end{tabular}
  \caption{Results for each frequency band to $1\sigma$.  Values for $g_{\rm NL}$, $\tau_{\rm NL}$ and $A_{\rm NL}$ on the top are for raw maps.  The values on the bottom are for foreground clean maps. \vspace{-0.3cm}}
  \label{tab:label}
\end{table}

After obtaining the theory, data and simulated curves we use the best fitting procedure described in~\cite{Smidt:2009ir} where we minimize $\chi^2$ to fit $\tau_{\rm NL}$ and $g_{\rm NL}$ simultaneously.  Our results are listed in Table~\ref{tab:label}.  We see that $g_{\rm NL}$ and $\tau_{\rm NL}$ are consistent with zero with 95\% confidence level ranges  $-7.4 < g_{\rm NL}/10^5 < 8.2$ and   $-0.6 < \tau_{\rm NL}/10^4 < 3.3$ for V+W-band in foreground-cleaned maps.  The 95\% confidence intervals of $g_{\rm NL}$ versus $\tau_{\rm NL}$ are plotted in Figure~\ref{fig:confidence} for each band.   Furthermore, we find that the constraint $ -3 < A_{\rm NL} < 21.4$.

{\it Conclusion.}---This paper gives the first direct constraints on $\tau_{\rm NL}$ and improves the previous indirect estimate by four orders of magnitude.  Our limit on $\tau_{\rm NL}$ is close to a level where interesting constraints can started to be placed on models available in the literature for primordial perturbations, such as due to cosmic strings\cite{Engel:2008fu}.

Furthermore, this paper is the first to constrain $A_{\rm NL}$.  This quantity is a model dependent relation between $\tau_{\rm NL}$ and $f_{\rm NL}$, that for some models can be used to further constrain the tensor-to-scaler-ratio. We uncover that that the data is consistent with  $A_{\rm NL}$ values between -3 and 21 at 95\% confidence.  Moreover, models with $A_{\rm NL}$ largely negitive, such as some arising from DBI inflation are disfavored~\cite{Engel:2008fu}.

\acknowledgments
We are grateful to Eiichiro Komatsu, and Kendrick Smith for assistance during
various stages of this work.  This work was supported by NSF CAREER AST-0645427 and NASA NNX10AD42G at UCI and STFC rolling grant ST/G002231/1 (DM).\vspace{-0.4cm}

\appendix

%
%

\end{document}